\documentclass[12pt]{article}

\input{symblibrary.tex}
\usepackage{graphicx} 

\newcommand\pubnumber{DPF2013-41}
\newcommand\pubdate{\today}

\def\napoli{Department of Physics and Astronomy\\
University of New Mexico, MSC07 4220, 1919 Lomas Blvd. NE,\\ Albuquerque, NM 87131-0001, USA}
\def\support{\footnote{Work supported by the U.S. Department of Energy.}}

\def\Title#1{\begin{center} {\Large #1 } \end{center}}
\def\Author#1{\begin{center}{ \sc #1} \end{center}}
\def\Address#1{\begin{center}{ \it #1} \end{center}}

\newcommand\pubblock{\rightline{\begin{tabular}{l} \pubnumber\\
         \pubdate  \end{tabular}}}
\newenvironment{Abstract}{\begin{quotation}  }{\end{quotation}}
\newenvironment{Presented}{\begin{quotation} \begin{center} 
             PRESENTED AT\end{center}\bigskip 
      \begin{center}\begin{large}}{\end{large}\end{center} \end{quotation}}
\def\Acknowledgments{\bigskip  \bigskip \begin{center} \begin{large}
             \bf ACKNOWLEDGMENTS \end{large}\end{center}}
\def\beq{\begin{equation}}
\def\eeq#1{\label{#1}\end{equation}}
\def\eeqn{\end{equation}}

\def\beqa{\begin{eqnarray}}
\def\eeqa#1{\label{#1}\end{eqnarray}}
\def\eeqan{\end{eqnarray}}

\let\bar=\overbar

\def\D{{\cal D}}

\def\Dslash{\not{\hbox{\kern-4pt $D$}}}
\def\dslash{\not{\hbox{\kern-2pt $\del$}}}

\def\BR{\mbox{\rm BR}}

\def\msb{{\bar{\ssstyle M \kern -1pt S}}}

\begin{document}
\begin{titlepage}
\pubblock

\vfill
\Title{Evidence for a bottom baryon resonance 
       \( \mathbf{\Lambda^{*0}_{b}} \) in CDF~data}
\vfill
\Author{ Prabhakar Palni (CDF Collaboration)\support}
\Address{\napoli}
\vfill
\begin{Abstract}
 Using data from proton-antiproton collisions at
      \(\sqrt{s}=1.96\tev\) recorded by the CDF~II~detector at the
      Fermilab Tevatron, evidence for the excited resonance state \(\LbS\)
      is presented in its \(\Lb\pim\pip\) decay, followed by the
      \(\Lb\to\Lc\pim\) and \(\Lc\to\pKpi\) decays.  The analysis is
      based on a data sample corresponding to an integrated luminosity
      of \({9.6}\invfb\) collected by an online event selection based on
      charged-particle tracks displaced from the proton-antiproton
      interaction point.  The significance of the observed signal is
      \(3.5\sigma \).  The mass of the observed state is found to be
      \(5919.22\,\pm0.76\,\mevcc\)
      in agreement with similar findings in proton-proton collision
      experiments.
\end{Abstract}
\vfill
\begin{Presented}
DPF 2013\\
The Meeting of the American Physical Society\\
Division of Particles and Fields\\
Santa Cruz, California, August 13--17, 2013\\
\end{Presented}
\vfill
\end{titlepage}
\def\thefootnote{\fnsymbol{footnote}}
\setcounter{footnote}{0}

\section{Introduction}
  Baryons with a heavy quark \( Q \) are useful for probing
  quantum chromodynamics (QCD) in its confinement domain.
  Observing new heavy-quark baryon states and measuring their
  properties provides further experimental constraints to the
  phenomenology in this regime.
  This report provides an additional contribution to the currently small
  number of heavy quark baryon observations.
\par  
  In the framework of heavy-quark effective theories
  (HQET)~\cite{Neubert:1993mb,Isgur:1989vq}, a bottom quark \(\b\) and a
  spin-zero \( [\u\d] \) diquark, carrying an angular momentum \(L=1\)
  relative to the \(\b\) quark (hence named \(P\)-wave states), can form
  two excited states.  These are named \LbS, with same quark content as
  the singlet \Lb~\cite{Beringer:1900zz} and isospin \(I=0\), but total
  spin and parity \(J^{P} = {\frac{1}{2}}^{-}\) and 
  \(J^{P} = {\frac{3}{2}}^{-}\)~\cite{Korner:1994nh}. 
  These isoscalar states are the lightest \(P\)-wave states that can
  decay to the \Lb baryon via strong interaction processes. The decays
  require the emission of a pair of low-momentum ({\it soft}) pions.
  Both \LbS~\cite{notation:lbs} particles are classified as
  bottom-baryon resonant states.
  Several recent theoretical predictions of their masses are 
  available.
  An approach based on a quark-potential model with the color hyperfine
  interaction is used in Ref.~\cite{Karliner:2008sv}. The authors in
  Ref.~\cite{Garcilazo:2007eh} use a constituent quark model
  incorporating the basic properties of QCD and solving exactly the
  three-body problem.
  A heavy baryon is considered in Ref.~\cite{Ebert:2005xj} as a
  heavy-quark and light-diquark system in the framework of the relativistic
  quark model based on the quasipotential approach in QCD. 
  The spectroscopy of isoscalar heavy baryons and their excitations is
  studied in Ref.~\cite{AzizaBaccouche:2001pu} within the framework of
  HQET at leading and next-to-leading orders in the combined inverse
  heavy-quark mass, \(1/m_{Q}\), and inverse number of colors,
  \(1/N_{c}\), expansions.
  The nonperturbative formalism of QCD sum rules is applied within
  HQET to calculate the mass spectra of the bottom baryon
  states~\cite{Zhang:2008pm}.
  Some calculations predict \LbS masses smaller than the hadronic
  decay kinematic-threshold (\(\approx5900\mevcc\)), allowing 
  only radiative decays~\cite{Garcilazo:2007eh,Zhang:2008pm}.
  Other calculations predict the mass difference
  \(M(\LbS)\,-\,M(\Lb)\) for the \(J^{P}={\frac{1}{2}}^{-}\) state to be
  approximately in the range
  300--310\(\mevcc\)~\cite{Karliner:2008sv,Ebert:2005xj,AzizaBaccouche:2001pu}.
  The mass splitting between the two states is predicted to be in the
  range 10--17\(\mevcc\).
\par 
  The first experimental studies of \(\b\)-quark baryon resonant states
  were reported by CDF with the observation of the \(S\)-wave states
  \Sgbst in their \(\Lb\pipm\) decays~\cite{:2007rw,CDF:2011ac}.
%
  %The first information on \(\b\)-quark baryon resonance was made by
  %CDF, which reported the \(S\)-wave \Sgbst states in their \(\Lb\pipm\)
  %decays~\cite{:2007rw} followed by their mass and width
  %measurements~\cite{CDF:2011ac}. 
%
  The ground states of the charged bottom-strange \(\Xi_{b}^{-}\)
  baryon~\cite{:2007ub,:2007un,Aaltonen:2009ny} and bottom
  doubly-strange \(\Omega_{b}^{-}\)~\cite{Abazov:2008qm,Aaltonen:2009ny}
  were reported by both CDF and \dzero, and later CDF observed the
  neutral bottom-strange baryon \(\Xi_{b}^{0}\)~\cite{Aaltonen:2011wd}.
  Recently, LHCb reported precise mass measurements of the ground state
  \(\Lb\), the \(\Xi_{b}^{-}\) state, and the
  \(\Omega_{b}^{-}\) state~\cite{Aaij:2013qja}.
  The CMS collaboration observed another bottom-strange state,
  \(\Xi_{b}^{*0}\), which is interpreted as a
  \(J^{P}={\frac{3}{2}}^{+}\) resonance~\cite{Chatrchyan:2012ni}. Most
  recently, two states, interpreted as the two \LbS resonant states were
  observed by the LHCb collaboration for the first
  time~\cite{Aaij:2012da}.
\par 
  In this report, we present evidence for the production of a \LbS
  resonance state in CDF data.  We search for candidate \LbS baryons
  produced in proton-antiproton collisions at \(\sqrt{s}=1.96\tev\)
  using a data sample from an integrated luminosity of \(9.6\invfb\)
  collected by CDF with a specialized online event-selection (trigger)
  that collects events enriched in fully hadronic decays of \(\b\)
  hadrons. The \LbS candidates are identified in the pseudorapidity
  range \(\rapid<1.0\) using their exclusive decays to \(\Lb\) baryons
  and two oppositely-charged soft pions.  The excellent performance of
  the CDF devices for measuring charged particle trajectories (tracks)
  allows reconstructing charged particles with transverse momenta as low
  as \(200\mevc \).  The result in this paper is the first to support
  the LHCb observation~\cite{Aaij:2012da}.
  
  \section{The CDF II Detector }
\par
  The component of the \(\cdf2 \) detector~\cite{Acosta:2004yw} most
  relevant to this analysis is the charged-particle tracking system,
  which operates in a uniform axial magnetic field of \(1.4\,{\rm T}\)
  generated by a superconducting solenoidal magnet.
  The inner tracking system is comprised of a silicon
  tracker~\cite{Aaltonen:2013uma}.  A large open-cell cylindrical drift
  chamber~\cite{Affolder:2003ep} completes the tracking system.
  The silicon tracking system measures the transverse impact parameter
  of tracks with respect to the primary interaction
  point, \({d_{0}}\)~\cite{cdf:coordinates}, with a resolution of
%
  %\({\sigma_{d_{0}}}\approx{45}\mkm\)~\cite{cdf:coordinates}, including
  %an approximately \({25}\mkm\) contribution from the beam size.
%
  \({\sigma_{d_{0}}}\approx{40}\mkm\), including
  an approximately \({32}\mkm\) contribution from the beam
  size~\cite{Aaltonen:2013uma}.
  The transverse momentum resolution of the tracking system is
  \({\sigma({\pt})}/{{\pt}^{2}}\approx{0.07\%}\) with \(\pt\) 
  in GeV/\(c\)~\cite{cdf:coordinates}.
\section{Data Sample and Simulation}  
\par 
  This analysis relies on a three-level trigger to collect data samples
  enriched in multibody hadronic decays of \b hadrons (displaced-track
  trigger).  The trigger requires two charged particles in the drift
  chamber, each with \(\pt>2.0\gevc\)~\cite{Thomson:2002xp}. The
  particle tracks are required to be azimuthally separated by
  \(2\degrees<\Delta\phi<90\degrees\)~\cite{cdf:coordinates}.  Silicon
  information is added and the impact parameter \({d_{0}}\) of each
  track is required to lie in the range 0.12--1\(\mm\) providing
  efficient discrimination of long-lived \b
  hadrons~\cite{Ashmanskas:2003gf}.  Finally, the distance \(\lxy\) in
  the transverse plane between the collision space-point (primary
  vertex) and the intersection point of the two tracks projected onto
  their total transverse momentum is required to exceed \(200\mum\).
\par  
  The mass resolution of the \LbS resonances is predicted with a
  Monte-Carlo simulation that generates \b quarks according to a
  calculation expanded at next-to-leading order in the strong coupling
  constant~\cite{Nason:1987xz} and produces events containing
  final-state hadrons by simulating \b-quark
  fragmentation~\cite{Peterson:1982ak}.
  In the simulations, the \LbS baryon is assigned the mass value of
  \(5920.0\mevcc\).
  %\(5920.0\mevcc\)~\cite{AzizaBaccouche:2001pu}.
%  
  Decays are simulated with the {\sc evtgen}~\cite{Lange:2001uf}
  program, and all \b hadrons are simulated unpolarized. 
  The generated events are passed to a {\sc geant3}-based~\cite{Geant}
  detector simulation, then to a trigger simulation, and finally the
  same reconstruction algorithm as used for experimental data.
%
  %The generated events are input to the detector and trigger simulation
  %based on {\sc geant3}~\cite{Geant} and processed through the same
  %reconstruction and analysis algorithms as are used on the data.
% 
%
%\section{Data sample and event selection}
%\label{sec:Data}
% 
\par 
  The \LbS candidates are reconstructed in the exclusive
  strong-interaction decay
  \(\LbS\to\Lb\pim_{\mathit{s}}\pip_{\mathit{s}}\), where 
  the low-momentum pions \(\pipm_{\mathit{s}}\) are produced near 
  kinematic threshold~\cite{notation:cc}.
  The \(\Lb\) baryon decays through the weak interaction to a baryon
  \(\Lc\) and a pion, labeled as \(\pim_{b}\) to distinguish it from the
  soft pions.  This is followed by the weak-interaction decay
  \(\Lc\to\pKpi\).
%  
  %followed by the weak-interaction decays \(\Lb\to\Lc\pim_{b}\) and
  %\(\Lc\to\pKpi\)~\cite{notation:cc}.
% 
%\subsection{Reconstruction of the \Lb candidates}
%\label{sec:reco-lb}
%
  We search for a \LbS signal in the \(Q\)-value distribution, where
    \(Q = m(\Lb\pim_{\mathit{s}}\pip_{\mathit{s}}) - m(\Lb) - 2\,{m_{\pi}}\,,\) 
  \(m(\Lb)\) is the reconstructed \(\Lc\pim_{b}\) mass, and \({m_{\pi}}\) 
  is the known charged-pion mass.  The effect of the \Lb mass resolution 
  is suppressed, and most of the systematic uncertainties are reduced in
  the mass difference.
  We search for narrow structures in 6--45\(\mevcc\) range of the
  \(Q\)-value spectrum motivated by the theoretical
  estimates~\cite{Karliner:2008sv,Ebert:2005xj,AzizaBaccouche:2001pu}
  and the LHCb findings~\cite{Aaij:2012da}.
\section{Analysis Overview} 
\par  
  The analysis begins with the reconstruction of the \(\LcpKpi\) decay
  space-point by fitting three tracks to a common point.
  Standard CDF quality requirements are applied to each track, and only
  tracks corresponding to particles with \(\pt>400\mevc \) are used.  No
  particle identification is used.  All tracks are refitted using pion,
  kaon, and proton mass hypotheses to correct for the mass-dependent
  effects of multiple scattering and ionization-energy loss.
  The invariant mass of the \(\Lc\) candidate is required to match the
  known value~\cite{Beringer:1900zz} within \(\pm18\mevcc \).
  The momentum vector of the \(\Lc\) candidate is then extrapolated to
  intersect with a fourth track that is assumed to be a pion, to form
  the \( \Lb\to\Lc\pim_{b} \) candidate. The \Lb
  reconstructed decay point (decay vertex) is subjected to a
  three-dimensional kinematic fit with the \(\Lc\) candidate mass
  constrained to its known value~\cite{Beringer:1900zz}. The probability
  of the \Lb vertex fit must exceed \(0.01\%\)~\cite{CDF:2011ac}.
  The proton from the \(\Lc\) candidate is required to have
  \(\pt>2.0\gevc \) to ensure that the proton is consistent with having
  contributed to the trigger decision.  The minimum requirement on
  \(\pt(\pim_{b})\) is determined by an optimization procedure,
  maximizing the quantity \(S_{\Lb}/(1+\sqrt{B})\)~\cite{Punzi:2003bu},
  where \(S_{\Lb}\) is the number of \(\Lb\) signal events obtained from
  the fit of the observed \(\Lc\pim_{b}\) mass distribution and
  \({B}\) is the number of events in the sideband region of
  \(50<Q<90\mevcc\) scaled to the background yield expected in the
  signal range \(14.0<Q<26.0\mevcc\).
%
% The form of the quantity takes into account the low background level.
%
  The sideband region boundaries are motivated by the signal predictions
  in Refs.~\cite{Karliner:2008sv,Ebert:2005xj,AzizaBaccouche:2001pu}.
  The resulting requirement is found to be \(\pt(\pim_{b})>1.0\gevc\).
%
  %The sideband spectrum is parametrized by a second order Chebyshev
  %polynomial.
%
  %The requirement of \(\pt(\pim_{b})>1.0\gevc\) corresponds
  %to the maximum of the score function. 
%
  The momentum criteria both for proton and \(\pim_{b}\) candidates
  favor these particles to be the two that contribute to the
  displaced-track trigger decision.
  To keep the soft pions from \LbS decays within the kinematic
  acceptance, the \Lb candidate must have \(\pt(\Lb)>9.0\gevc \). This
  maximizes the quantity \( S_{\rm MC}/(1+\sqrt{B}) \), where 
  \(S_{\rm{MC}}\) is the \(\LbS \) signal reconstructed in the simulation.
%
  %and \({B}\) is the number of events in the previously defined sideband
  %region of the \(\LbS \) \(Q\)-value spectrum.
%
\par 
  To suppress prompt backgrounds from primary interactions, the decay
  vertex of the long-lived \Lb candidate is required to be distinct from
  the primary vertex by requiring the proper decay time and its
  significance to be \(\ct{(\Lb)}>200\mkm \) and
  \(\ct(\Lb)/\sigma_{\it{ct}}>6.0 \), respectively. The first criterion
  validates the trigger condition, while the second is fully efficient
  on simulated \LbS signal decays.  We define the proper decay time as
     \( \ct{(\Lb)} = \lxy\,{{m_{\Lb}}\,{c}}/{\pt}\,\,,\) 
  where \({m_{\Lb}} \) is the known mass of the \Lb
  baryon~\cite{Beringer:1900zz}.
%
  %For \(\lxy\) we use the best primary
  %vertex determined for the event. 
%
  %The primary vertex is determined 
  %event-by-event when computing this vertex displacement.
%  
  We require the \Lc vertex to be associated with a \Lb decay by
  requiring \(\ct(\Lc)>-100\,\mkm \), as derived from the quantity
  \( \lxy(\Lc) \) measured with respect to the \Lb vertex. This
  requirement reduces contributions from \Lc baryons directly produced
  in \(\proton\antiproton\) interactions and from random combinations
  of tracks that accidentally are reconstructed as \Lc candidates.
%
  %(which may have negative \(\ct(\Lc)\) values).
% 
  To reduce combinatorial background and contributions from
  partially-reconstructed decays, \Lb candidates are required to point
  towards the primary vertex by requiring the impact parameter
  \({{d_{0}}(\Lb)}\) not to exceed \( 80\,\mkm \). The \(\ct(\Lc)\) and
  \({{d_{0}}(\Lb)}\) criteria~\cite{CDF:2011ac} are fully efficient for
  the \LbS signal.
\par
\begin{figure}
\begin{center}
  \includegraphics[width=0.49\textwidth]
% {figures/lamb.tot9600pb.503.2nd-cdfdraft.eps}
  {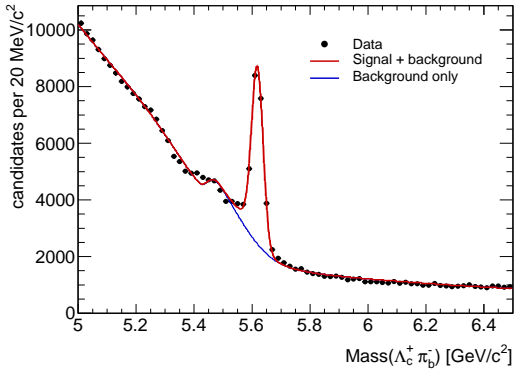}
\caption{ Invariant mass distribution of \(\Lb\to\Lc\pim_{b}\)
          candidates with a fit overlaid. The shoulder at the left 
          sideband is dominated by fully reconstructed \B mesons and
          partially reconstructed \Lb decays.
          \label{fig:signal-lb} }
\end{center}
\end{figure}

 
  Figure~\ref{fig:signal-lb} shows the resulting prominent \Lb signal in
  the \(\Lc\pim_{b}\) invariant mass distribution.
%
  %A fit is applied to the data using
  %a model that comprises a Gaussian \(\Lb\to\Lc\pim_{b}\) signal on a
  %background shaped by several
  %contributions~\cite{:2007rw,CDF:2011ac,Abulencia:2006df}. 
%
  The binned maximum-likelihood fit finds a signal of approximately
  \(15\,400 \) candidates at the expected \Lb mass, with unity
  signal-to-background ratio.
  The fit model describing the invariant mass distribution comprises the
  Gaussian \(\Lb\to\Lc\pim_{b}\) signal overlapping a background shaped by
  several contributions. Random four-track combinations dominating the
  right sideband are modeled with an exponentially decreasing function.
  Coherent sources populate the left sideband and leak under the
  signal. These include reconstructed \B mesons that pass the
  \(\Lb\to\Lc\pim_{b}\) selection criteria, partially reconstructed \Lb
  decays, and fully reconstructed \Lb decays other than \(\Lc\pim_{b}\)
  (\exegrat,~\(\Lb\to\Lc\Km\)). Shapes representing the physical
  background sources are derived from Monte-Carlo simulations.  Their
  normalizations are constrained to branching ratios that are either
  measured (for \B meson decays, reconstructed within the same
  \(\Lc\pim_{b}\) sample) or theoretically predicted (for \Lb
  decays). 
  The discrepancy between the fit and the data at smaller masses than
  the \Lb signal is attributed to incomplete knowledge of the branching
  fractions of decays populating this
  region~\cite{Abulencia:2006df,:2007rw,Aaltonen:2009zn,CDF:2011ac} and
  is verified to have no effect on the final results.
%
  %The fit is used only to define the size of the signal extent
  %at the fitted \Lb mass.
  %The fit is used only to define the signal region around the fitted \Lb
  %mass.
  The fit is used only to define the \LbS search sample.
% 
%  
%\subsection{Reconstruction of \LbS candidates}
%\label{sec:reco-sgb}
\section{Experimental Mass Distribution and The Fit }
\par 
  To reconstruct the \(\LbS\) candidates, each \(\Lb\) candidate with
  mass within the range 5.561--5.677\(\gevcc\)
  (\(\pm3\sigma\)) is combined with a pair of oppositely-charged
  particles, each assigned the pion mass.
  To increase the efficiency for reconstructing \LbS decays near the
  kinematic threshold, the quality criteria applied to soft-pion tracks
  are loosened. The basic requirements for hits in the drift chamber and
  main silicon tracker are imposed on the \(\pipm_{\mathit{s}}\) tracks,
  and tracks reconstructed with a valid fit, proper error matrix, and
  with \(\pt>200\mevc\) are accepted.
%
% having hits in both detectors 
%
  The relaxed requirements on the soft-pion tracks increase the
  reconstructed \LbS candidates yield by a factor of approximately
  \(2.6\).
  %more \LbS candidates.
% 
\par  
  To reduce the background, a kinematic fit is applied to the
  resulting \(\Lb\), \(\pim_{\mathit{s}}\) and \(\pip_{\mathit{s}}\)
  candidates that constrains them to originate from a common point. The
  \(\Lb\) candidates are not constrained to the \(\Lb\) mass in
  this fit.  Furthermore, since the bottom-baryon resonance originates
  and decays at the primary vertex, the soft-pion tracks are required to
  originate from the primary vertex by requiring an impact parameter
  significance
   \( {{d_{0}}(\pipm_{\mathit{s}})/{\sigma_{d_{0}}}} \) 
  smaller than \(3\)~\cite{:2007rw,CDF:2011ac}, determined by maximizing
  the quantity \( S_{\rm MC}/(1+\sqrt{B})\).
%
%\section{Experimental mass distribution and the fit}
%\label{sec:signal}
% 
\section{Results and Conclusions}  
\par
\begin{figure}   
\begin{center}  
  \includegraphics[width=0.49\textwidth] 
%  {figures/lbs.run-55.2nd-cdfdraft.eps}  
   {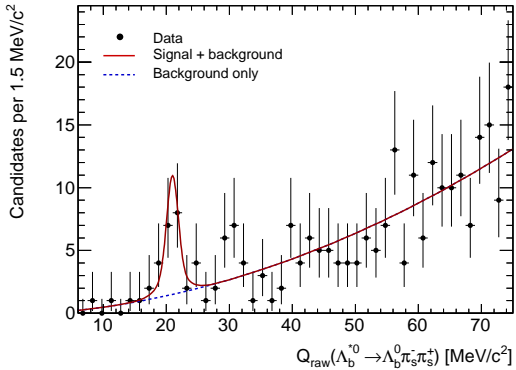}  
  \caption{ Distribution of \(Q\)-value for \LbS candidates, with fit
            projection overlaid. }
\label{fig:signal-lbs}
\end{center}
\end{figure}

  
  The observed \(Q\)-value distribution is shown in
  Fig.~\ref{fig:signal-lbs}.  A narrow structure at \(Q\approx21\mevcc\)
  is clearly seen.
  The projection of the corresponding unbinned likelihood fit is
  overlaid on the data.  The fit function includes a signal and a smooth
  background.  The signal is parametrized by two Gaussian functions with
  common mean, and widths and relative sizes set according to
  Monte-Carlo simulation studies.  Approximately \(70\% \) of the signal
  function is a narrow core with \(0.9\mevcc\) width, while the wider
  tail portion has a width of about \(2.3\mevcc\).  The background is
  described by a second order polynomial.  The fit parameters are the
  position of the signal and its event yield.
  The negative logarithm of the extended likelihood function is
  minimized over the unbinned set of \(Q\)-values observed.
  The fit over the \(Q\) range 6--75\(\mevcc\) finds
  \(17.3^{+5.3}_{-4.6}\) signal candidates at 
  \(Q={20.96}\pm{0.35}\,\mevcc\).
\par 
  The significance of the signal is determined using a
  \({\log}\)-likelihood-ratio statistic,
  \(D=-2\ln({\mathcal{L}_0}/{\mathcal{L}_1})\)~\cite{Wilks,Royall}.
  We define the hypothesis \({\mathcal{H}_1}\) as corresponding to the
  presence of a \( \LbS \) signal in addition to the background and
  described by the likelihood \({\mathcal{L}_1}\).  The null hypothesis,
  \({\mathcal{H}_0}\), assumes the presence of only background with a
  mass distribution described by the likelihood \({\mathcal{L}_0}\), and
  is nested in \({\mathcal{H}_1}\).  The \({\mathcal{H}_1}\) hypothesis
  involves two additional degrees of freedom with respect to
  \({\mathcal{H}_0}\), the signal position and its size.
  The significance for a \(Q \) search-window of 6--45\(\mevcc\) is
  determined by evaluating the distribution of the \({\log}\)-likelihood
  ratio in pseudoexperiments simulated under the \({\mathcal{H}_0}\)
  hypothesis.  The fraction of the generated trials yielding a value of
  \(D\) larger than that observed in experimental data determines the
  significance. The fraction is \(2.3\times10^{-4}\), corresponding to a
  significance for the signal equivalent to \(3.5\) one-tailed Gaussian
  standard deviations.
%
%  
%\section{Systematic uncertainties}
%\label{sec:sys} 
%
\par   

  The systematic uncertainties on the mass determination derive from the
  tracker momentum scale, the resolution model, and the choice of the
  background model.  To calibrate the momentum scale, the energy loss in
  the tracker material and the intensity of the magnetic field must be
  determined.  Both effects are calibrated and analyzed in detail using
  large samples of \( \jpsi \), \( \psitwos \), \(\OneS\), and \(\Z\)
  particles reconstructed in the \(\mumu \) decay modes as well as
  \(\Dstarp\to\Dz(\to\Km\pip)\pip\), and
  \(\psitwos\to\jpsi(\to\mumu)\pipi\)
  samples~\cite{Acosta:2005mq,Aaltonen:2009vj}.  The corresponding
  corrections are taken into account by the tracking algorithms.  Any
  systematic uncertainties on these corrections are negligible in the
  \(Q\)-value measurements due to the mass difference term,
  \(m(\Lb\pim_{\mathit{s}}\pip_{\mathit{s}})-m(\Lb)\).
  The uncertainties on the measured mass differences due to the momentum
  scale of the low-\(\pt \) \({\pipm_{\mathit{s}}} \) tracks are
  estimated from a large calibration sample of
  \(\Dstarp\to\Dz\pip_{\mathit{s}} \) decays.  A scale factor of
  \(0.990\pm0.001\) for the soft pion transverse momentum is found to
  correct the difference between the \(Q \)-value observed in
  \(\Dstarp\) decays and its known value~\cite{Beringer:1900zz}. The
  same factor applied to the soft pions in a full simulation of
  \(\LbS\to\Lb\pim_{\mathit{s}}\pip_{\mathit{s}} \) decays yields a
  \(Q\)-value change of \(-0.28\mevcc \). Taking the full value of the
  change as the uncertainty, we adjust the \(Q \)-value determined by
  the fit to the \(\LbS \) candidates by \(-0.28\pm0.28\,\mevcc\).
  The Monte-Carlo simulation underestimates the detector resolution, and
  the uncertainty of this mismatch is considered as another source of 
  systematic uncertainty~\cite{CDF:2011ac}.
  To evaluate the systematic uncertainty due to the resolution, we use a
  model with floating width parameter where only the ratio of the widths
  of the two Gaussians is fixed. The resulting uncertainty is found to
  be \(\pm0.11\,\mevcc\).
  To estimate the uncertainty associated with the choice of background
  shape, we increase the degree of the chosen polynomial and find the
  uncertainty to be \(\pm0.03\,\mevcc\).  The statistical uncertainties
  on the resolution-model parameters due to the finite size of the
  simulated data sets introduce a negligible contribution.  Adding in
  quadrature the uncertainties of all sources results in a total 
  \(Q\)-value systematic uncertainty of \(\pm0.30\,\mevcc\).
%  
%  
%\section{Results and conclusions}
%\label{sec:results}
%
\par  
  Hence, the measured \(Q\)-value of the identified \LbS state is
  found to be
  \(20.68\,\pm\,0.35\stat\,\pm\,0.30\syst\,\mevcc \) \cite{Aaltonen:2013tta}.
  Using the known values of the charged pion and \Lb baryon
  masses~\cite{Beringer:1900zz}, we obtain the absolute \LbS mass value
  to be
  \(5919.22\,\pm0.35\,\stat\pm0.30\,\syst\pm0.60\,({\Lb})\,\mevcc\), 
  where the last uncertainty is the world's average \Lb mass 
  uncertainty reported in Ref.~\cite{Beringer:1900zz}.
%
%  with its uncertainty contributing to the
%  systematic uncertainty  on the \LbS absolute mass.
% 
  The result is closest to the calculation based on \(1/m_{Q},\,1/N_{c}\)
  expansions~\cite{AzizaBaccouche:2001pu}.
  The result is also consistent with the higher state \(\LbS(5920)\)
  recently observed by the LHCb experiment~\cite{Aaij:2012da}.
  LHCb also reports a state at approximately
  5912\(\mevcc\)~\cite{Aaij:2012da}.  Assuming similar relative
  production rates and relative efficiencies for reconstructing the
  \(\LbS(5912)\) and \(\LbS(5920)\) states in the CDF~II and LHCb
  detectors, the lack of a visible \(\LbS(5912)\) signal in our data is
  statistically consistent
%
%  (\(p\){-value}~\(=0.03\)) 
   within \(2\sigma\)
  with the \(\LbS(5912)\) yield reported by LHCb.
\par 
  In conclusion, we conduct a search for the \(\LbS\to\Lb\pim\pip\)
  resonance state in its \(Q\)-value spectrum.  A narrow structure is
  identified at \(5919.22\,\pm0.76\,\mevcc\) mass with a significance of
  \(3.5\sigma\).
  This signal is attributed to the orbital excitation of the bottom
  baryon \(\Lb\) and supports similar findings in proton-proton
  collisions.
\Acknowledgments
I would like to thank the U.S. Department of Energy for supporting this work.


\begin{thebibliography}{99}
\bibitem{cdf:coordinates}
   We use a cylindrical coordinate system with
  \(z\) axis along the nominal proton beam line, radius \(r\) measured
  from the beam line and \(\phi\) defined as an azimuthal angle. The
  transverse plane \((r,\phi)\) is perpendicular to the \(z\) axis.  The
  polar angle \(\theta\) is measured from the \(z\) axis. Transverse
  momentum \(\pt\) is the component of the particle's momentum projected
  onto the transverse plane.  Pseudorapidity is defined as
  \(\eta\,\equiv\,-\ln(\tan(\theta/2))\).  The impact parameter of a
  charged particle track \(d_{0}\) is defined as the distance of closest
  approach of the particle track to the point of origin (primary vertex)
  in the transverse  plane.
\end{thebibliography}
\end{document}